**Structure maps for MAX phases formability revisited**

Yiming Zhang[a,*], Yongjia Xu[a], Qing Huang[a], Shiyu Du[a], Mian Li[a], Youbing Li[a], Zeyu Mao[b], Qi Han[c]

[a] *Engineering Laboratory of Advanced Energy Materials, Qianwan Institute of CNiTECH, Ningbo, China*

[b] *State Key Laboratory for Mechanical Behavior of Materials, Xi'an Jiaotong University, Xi'an, China*

[c] *College of Science and Technology, Ningbo University, Ningbo, China*

## Abstract

The extraordinary chemical diversity of MAX phases raises the question of how many and which novel ones are yet to be discovered. The conventional schemes rely either on executions of well designed experiments or elaborately crafted calculations; both of which have been key tactics within the past several decades that have yielded many of important new materials we are studying and using today. However, these approaches are expensive despite the emergence of high throughput automations or evolution of high-speed computers. In this work, we have revisited the *in prior* proposed "light-duty" strategy, *i.e.* structure mapping, for describing the genomic conditions under which one MAX phase could form; that allow us to make successful formability/non-formability separation of MAX phases with a fidelity of 95.5%. Our results suggest that the proposed coordinates, and further the developed structure maps, are able to offer a useful initial guiding principles for systematic screenings of potential MAX phases and provide untapped opportunities for their structure prediction and materials design.

---

[*] Corresponding author

*E-mail address*: ymzhang@nimte.ac.cn (Y. Zhang)

## 1. Introduction

The MAX phases, or $M_{n+1}AX_n$, are a family of layered compounds within the class of complex carbides, nitrides, or borides; where "M" refers to an early transition metal, "A" belongs to an A-group element, "X" can be C, N and/or B, and "*n*" is an integer (commonly 1, 2, or 3) [1-5]. Based on the values of integer *n*, the MAX phases can be subcategorized into 211, 312 and 413 structures, respectively. Fig. 1(a) illustrates the hexagonal unit cells of 211, 312 and 413 MAX phases, which consist of $M_6X$ octahedra interleaved with layers of A elements. Currently, this kind of material is keeping attractive due to: **1)** their owning of unexpected combination of both metallic and ceramic properties, **2)** the ease by which one can regulate their chemistry, while keeping the structures same, **3)** the discovery of both in- and out-of-plane ordered phases that opens the door to broad their family members, and **4)** the ease by which they can be exfoliated into their 2D counterparts, MXenes, which exhibit fascinating properties and possess high application potentials to address some of society's most pressing issues [6]. At the moment, there are around 110 ternary MAX phases been experimentally reported, and new compounds with possible M, A and X elements are still being discovered [7-17]. Further, this number increases considerably when sold solutions or medium-/high-entropy candidates are taken into accounts [7, 18-22] (shown in Fig. 1(b)). The combinatorial design space suggests that the currently known MAX phases are only the tip of the iceberg, and substantial number of hitherto-unknown candidates are awaiting to be explored. During the past decade, *ab initio* calculations have been extensively conducted; either through estimating whether a compound is stable on an absolute scale, or demonstrating that a given MAX phase is more stable than all other competing phases, to investigate the potential MAX phases [23-34]. The predictive power and accuracy of these tools are evident; yet without time-consuming full DFT calculations, these analyses incapable of revealing any correlations or systematic behaviors [2, 23]. Accordingly, the extraordinary chemical diversity of MAX phases raises the issue of proposing general chemical design rules to tackle the factorial complexity of MAX phases design without arduous number crunching.

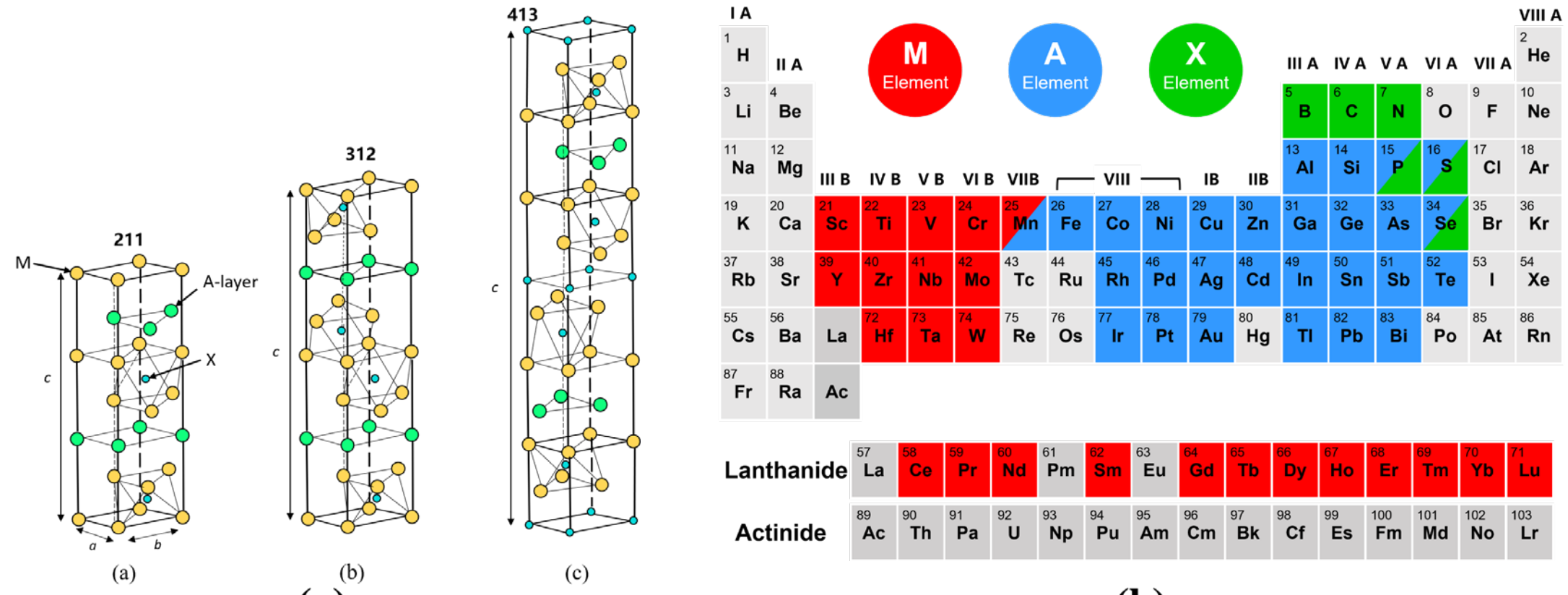

**(a)** **(b)**

Fig. 1 (a) Crystal structures for the 211, 312 and 413 MAX phases; (b) The introduced elements within various MAX phases.

Structure mapping, a scheme tries to discern key parameters that are strongly associated with the occurrence of a given crystal chemistry, has played an important role as a useful *a priori* guide for mapping out certain crystal types and served as a visualization tool for composition-structure relationships in a bivariate way [35-42]. There is a time-honored and distinguished tradition of such maps for novel materials developments [43-58]; and currently, it has been acknowledged that this creative and insightful graphical representations of data could leverage ones to grasp most important points without laborious analysis [59, 60]. In 2020, this strategy has been implemented to develop a classification scheme for MAX phases formability/non-formability separation [6]. Currently, the buzzing characteristic of novel composition development has enriched the family members of MAX phases; and the proposed structure map need to be validated and updated against the whole members of currently known MAX phases. In this work we endeavor to propose the revisited genomic conditions, that is composing of *geometric* and *electron concentration* factors, for deriving upgraded structure maps to identify the stability region that the $M_{n+1}AX_n$ phases could form, which can be further adopted to guide the exploration of potential MAX phases.

## 2. Revisiting the factors regulating MAX phases

In previous work [6], the important roles of *geometrical* and *electron concentration* factors play within the formability of MAX phases are fully discussed; and the expressions of these two factors are derived accordingly (shown in Eqns. (1) and (2)).

$$\textit{Geometrical Factor}: \frac{|R_M - R_A|}{R_M} \quad (1)$$

Where $R_M$ and $R_A$ represent atomic radii of M-site elements and A-site elements respectively, whose values are taken from CRC Handbook [61].

$$\textit{Electron Concentration Factor}: \frac{(VEC)_M * n_M + \left(\frac{e}{a}\right)_A * n_A + (VEC)_X * n_X}{n_M + n_A + n_X} \quad (2)$$

Where $(VEC)_M$ represents the number of valence electrons of M-site elements; $\left(\frac{e}{a}\right)_A$ represents the number of itinerant electrons of A-site elements; $(VEC)_X$ represents the number of valence electrons of X-site elements; and $n_M$, $n_A$, $n_X$ represent atomic coefficients of M-, A- and X- site elements respectively. Here, it worth noting that, although both play crucial roles for the chemical bonding, the *e/a* is introduced mainly for metallic bonding; whereas the *VEC* is defined locally for strong chemical bonding. Due to the difference, *VEC* takes the *d*-electrons accommodated in the valence band into account [62-65]. The values of valence electrons are collected from general inorganic chemistry textbook [66]; and the itinerant electrons are taken from Mizutani and Sato [63] (Table 1 lists the valence/itinerant electron values adopted in this work).

It can be noticed that the *electron concentration factor* in that study surveys the influences from M-, A- and X- site elements, as well as the stoichiometry factor, *n*; and the results have shown its feasibility. Therefrom the *electron concentration factor* is inherited into current research. In contrast, for the *geometrical* factor, only the size difference between M-site and A-site elements are under consideration; where the X-site elements dimension, as well as different *n* values, is disregarded. Owing to this, the following section is devoted to revisiting the *geometrical factor* extensively.

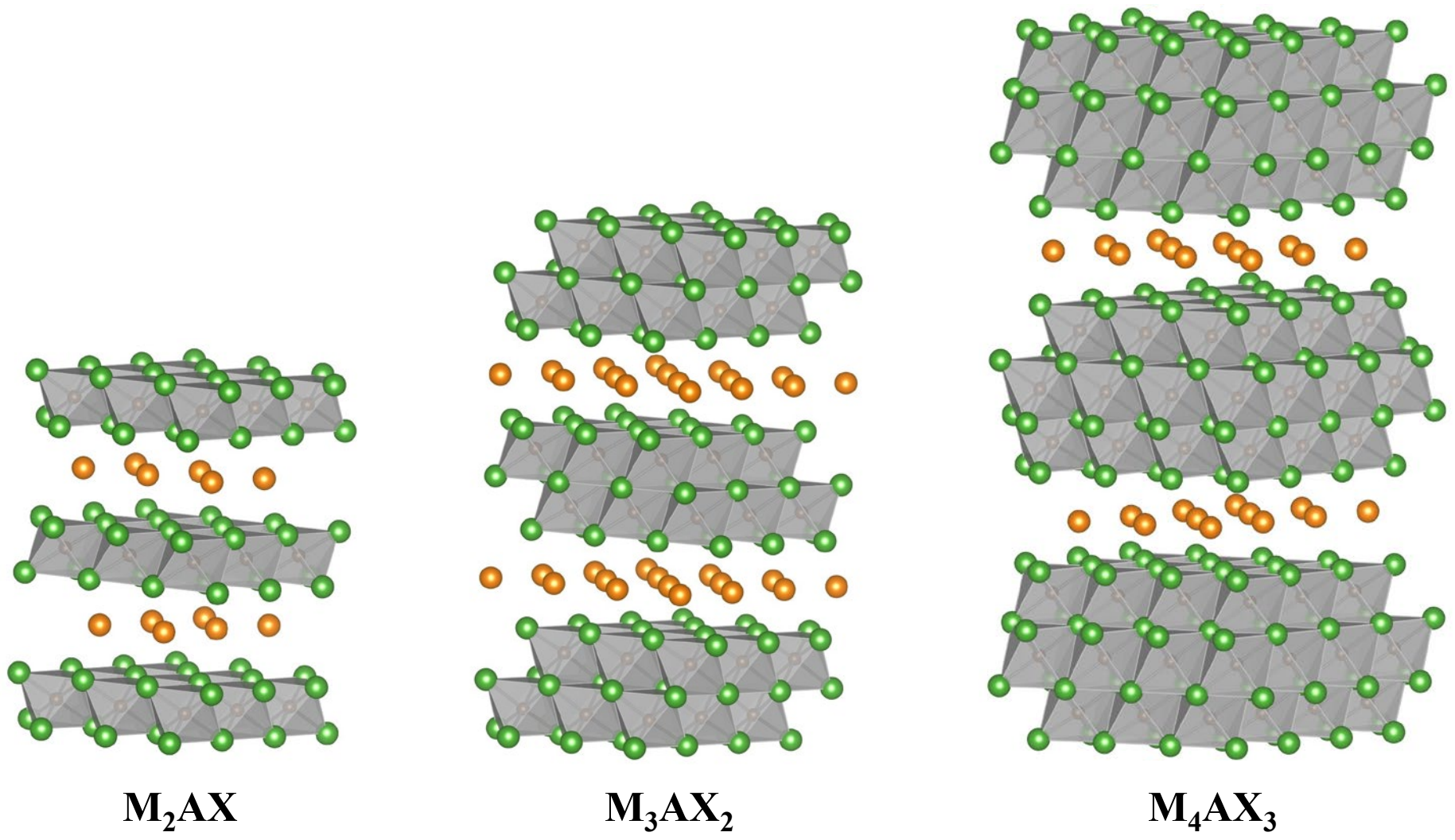

Fig. 2 Schematic illustration of the 211, 312, and 413 MAX phases crystal structure.

### 2.1 Development of *intercalation factor*

As one type of Nowotny octahedral phases, 211-MAX phases possess the crystal structure in which the $M_6X$-octahedra share common edges, with A-site elements intercalated [3, 67]. Fig. 2 depicts the crystal structures of 211, 312, and 413 MAX phases, where the difference between these three structures is in the numbers of octahedra block layers. By borrowing the idea of employing the size difference to characterize the geometrical mismatch [6, 68-70], the difference between the heights of $M_6X$ octahedra/octahedron and atomic radius of A-site elements (*i.e.*, the distance between opposite faces) is adopted, as one tolerance factor (called *intercalation factor* in this work), to gauge the geometrical mismatch of $M_6X$ block and A-site element. For the octahedra with side length of $a$, it can be shown that the height $h = \frac{\sqrt{6}}{3}a$; and hence the size difference between $M_6X$ octahedra and A-site element can be expressed as:

$$\frac{\left|\frac{2}{3}\sqrt{6}R_M - R_A\right|}{\frac{2}{3}\sqrt{6}R_M} \quad (3)$$

Where $R_M$ and $R_A$ represent atomic radii of M-site elements and A-site elements respectively.

As shown in Fig. 2, there are $n$ layers of $M_6X$ octahedron separating each A- layer for $M_{n+1}AX_n$ phases; that is, in 211 phases there is one $M_6X$ octahedron layer; in 312 phases two $M_6X$ octahedron layers, and in 413 phases three $M_6X$ octahedron layers. Thus, the *intercalation factor* shown in Eq. (3) for 312 and 413 phases can be rewritten as $\frac{\left|2*\frac{2}{3}\sqrt{6}R_M - R_A\right|}{2*\frac{2}{3}\sqrt{6}R_M}$ and $\frac{\left|3*\frac{2}{3}\sqrt{6}R_M - R_A\right|}{3*\frac{2}{3}\sqrt{6}R_M}$, respectively. In general, the *intercalation factor* for $M_{n+1}AX_n$ phases can be expressed as:

$$\frac{\left|n*\frac{2}{3}\sqrt{6}R_M - R_A\right|}{n*\frac{2}{3}\sqrt{6}R_M} \quad (4)$$

## 2.2 Introduction of *octahedral factor*

Further, by drawing lessens from tolerant factors designed for perovskites (*i.e.* another important type of Nowotny octahedral phases, but is a kind of valence compound), the *octahedral factor* is introduced in as the other tolerance factor [71-74]. For MAX phases, this is represented as:

$$\frac{R_X}{R_M} \quad (5)$$

Where $R_X$ and $R_M$ represent atomic radii of X-site elements and M-site elements respectively.

The *intercalation factor* and *octahedral factor*, shown in Eqs. (4) and (5), constitute the components of *new geometrical factor* (shown in Eq. (6)); which takes the sizes of M-, A-, X-sites elements and $M_6X$ blocks, as well as the effect of stoichiometry factor, $n$, into consideration. Here, it worth noting that the geometrical factor is for characterization of the lattice distortion, so that for alloys a linear combination of the sizes, with their stoichiometric ratios, of the constituent elements on M-, A- or X-site are employed.

$$\textit{New Geometrical Factor:}\ \frac{\left|n*\frac{2}{3}\sqrt{6}R_M-R_A\right|}{n*\frac{2}{3}\sqrt{6}R_M}+\frac{R_X}{R_M} \qquad (6)$$

The values of atomic radii are taken from CRC Handbook [61]; and Table 2 listed the atomic radius values adopted in this work. In the next part, the *electron concentration factor* (shown in Eq. (2)) and *new geometrical factor* (shown in Eq. (6)) are employed to construct structure maps for framing out the formable regions of $M_{n+1}AX_n$ phases.

**Table 1 The valence/itinerant electron values adopted in this work** [63, 66].

| M-site | | A-site | | X-site | |
|---|---|---|---|---|---|
| Elements | Valence Electrons | Elements | Itinerant Electrons | Elements | Valence Electrons |
| Sc | 3 | Al | 3.01 | C | 4 |
| Ti | 4 | Si | 4.00 | N | 5 |
| V | 5 | P | 4.97 | B | 3 |
| Cr | 6 | S** | 6.00 | P | 5 |
| Mn | 7 | Mn | 1.05 | S | 6 |
| Y | 3 | Fe | 1.05 | Se | 6 |
| Zr | 4 | Co | 1.03 | | |
| Nb | 5 | Ni | 1.16 | | |
| Mo | 6 | Cu | 1.00 | | |
| Hf | 4 | Zn | 2.04 | | |
| Ta | 5 | Ga | 3.00 | | |
| W | 6 | Ge | 4.05 | | |
| Ce* | 3 | As | 4.92 | | |
| Pr | 2 | Se** | 6.00 | | |
| Nd | 2 | Rh | 1.00 | | |
| Sm | 2 | Pd | 0.96 | | |
| Gd | 3 | Ag | 1.01 | | |
| Tb | 2 | Cd | 2.03 | | |
| Dy | 2 | In | 3.03 | | |
| Ho | 2 | Sn | 3.97 | | |
| Er | 2 | Sb | 4.99 | | |
| Tm | 2 | Te** | 6.00 | | |
| Yb | 2 | Ir | 1.60 | | |
| Lu | 3 | Pt | 1.63 | | |
| | | Au | 1.00 | | |
| | | Tl | 3.03 | | |
| | | Pb | 4.00 | | |
| | | Bi | 4.94 | | |

* Due to the 4*f* orbitals penetrate the xenon core appreciably, they cannot overlap with ligand orbitals; and so, do not participate significantly in bonding [75]. Therefore, the 4*f* electrons of lanthanides are not counted.

** Valence electrons are used

**Table 2 The atomic radius values adopted in this work (Unit: Å)** [61].

| M-site | | A-site | | X-site | |
|---|---|---|---|---|---|
| Elements | Atomic Radius | Elements | Atomic Radius | Elements | Atomic Radius |
| Sc | 1.59 | Al | 1.24 | C | 0.75 |
| Ti | 1.48 | Si | 1.14 | N | 0.71 |
| V | 1.44 | P | 1.09 | B | 0.84 |
| Cr | 1.30 | S | 1.04 | P | 1.09 |
| Mn | 1.29 | Mn | 1.29 | S | 1.04 |
| Y | 1.76 | Fe | 1.24 | Se | 1.18 |
| Zr | 1.64 | Co | 1.18 | | |
| Nb | 1.56 | Ni | 1.17 | | |
| Mo | 1.46 | Cu | 1.22 | | |
| Hf | 1.64 | Zn | 1.20 | | |
| Ta | 1.58 | Ga | 1.23 | | |
| W | 1.50 | Ge | 1.20 | | |
| Ce | 1.84 | As | 1.20 | | |
| Pr | 1.90 | Se | 1.18 | | |
| Nd | 1.88 | Rh | 1.34 | | |
| Sm | 1.85 | Pd | 1.30 | | |
| Gd | 1.82 | Ag | 1.36 | | |
| Tb | 1.81 | Cd | 1.40 | | |
| Dy | 1.80 | In | 1.42 | | |
| Ho | 1.79 | Sn | 1.40 | | |
| Er | 1.77 | Sb | 1.40 | | |
| Tm | 1.77 | Te | 1.37 | | |
| Yb | 1.78 | Ir | 1.32 | | |
| Lu | 1.74 | Pt | 1.30 | | |
| | | Au | 1.30 | | |
| | | Tl | 1.44 | | |
| | | Pb | 1.45 | | |
| | | Bi | 1.50 | | |

## 3 Data collection

For constructing structure maps, the data of formable and non-formable MAX phases are collected. In this work, comprehensive but not exhaustive list of the experimental synthesized MAX phases (*i.e.* hypothetical MAX phases, like $Sc_2AlC$, $Y_2AlC$ and $Mo_2AlC$, are not included), which have definite compositions, are recorded as formable MAX phases (a total of 202 systems, shown in Table 3 and 4); and the ones that are screened based on elastic and thermodynamic stability by using *ab initio* calculations are recorded as non-formable MAX phases (a total of 92 systems, shown in Table 5) [26]. Here, the incentives of the choice for non-formable MAX phases are worth to be emphasized as follow.

It might be questioned that the non-formable phases are chosen based on the investigations without full stability calculations. It is acknowledged that there are valuable works that systematically investigate the stabilities of MAX phases *via* showing one given phase is more stable than all other competing phases [25, 29-31, 33], which brings a more rigorous criterion to discern possible MAX phases; and when done it properly, could screen potential MAX phases out showing excellent agreement with experimental observed ones [2]. Nevertheless, novel preparation strategies have recently been developed to synthesis MAX phases successfully *via* avoiding the reactions to form competitive phases [8, 10, 14, 76, 77]; and the adoptions of unstable phases derived from full stability computations would obstruct the classifications. In contrast, the two criteria employed by Aryal *et al.* [26] are necessary but not sufficient conditions for MAX phase formability; and the non-formable ones screened out by following these two criteria are the ones that would scarcely form.

**Table 3 Comprehensive list of experimentally synthesized ternary MAX Phases[a]**

| Al | Si | P | S |
|---|---|---|---|
| $Ti_2AlC$<br>$V_2AlC$<br>$Cr_2AlC$<br>$Nb_2AlC$<br>$Ta_2AlC$<br>$Ti_2AlN$<br>$Zr_2AlC$<br>$Hf_2AlC$<br>$Ti_3AlC_2$<br>$Ta_3AlC_2$<br>$Zr_3AlC_2$<br>$Hf_3AlC_2$<br>$Ti_4AlN_3$<br>$Ta_4AlC_3$<br>$Nb_4AlC_3$<br>$V_4AlC_3$ | $Ti_3SiC_2$<br>$Ti_4SiC_3$ | $V_2PC$<br>$Nb_2PC$ | $Ti_2SC$<br>$Zr_2SC$<br>$Nb_2SC$<br>$Hf_2SC$<br>$Zr_2SB$<br>$Hf_2SB$<br>$Nb_2SB$ |
| **Ga** | **Ge** | **As** | **Se** |
| $Ti_2GaC$<br>$V_2GaC$<br>$Cr_2GaC$<br>$Nb_2GaC$<br>$Mo_2GaC$<br>$Ta_2GaC$<br>$Mn_2GaC$<br>$Ti_2GaN$<br>$V_2GaN$<br>$Sc_2GaC$<br>$Cr_2GaN$<br>$Ti_3GaC_2$<br>$Ta_4GaC_3$<br>$Ti_4GaC_3$ | $Ti_2GeC$<br>$V_2GeC$<br>$Cr_2GeC$<br>$Nb_2GeC$<br>$Ti_3GeC_2$<br>$Ti_4GeC_3$ | $V_2AsC$<br>$Nb_2AsC$ | $Zr_2SeC$<br>$Hf_2SeC$<br>$Zr_2SeB$<br>$Hf_2SeB$<br>$Zr_2SeP$<br>$Hf_2SeP$ |
| **In** | **Sn** | **Sb** | **Te** |
| $Sc_2InC$<br>$Ti_2InC$<br>$Zr_2InC$<br>$Nb_2InC$<br>$Hf_2InC$<br>$Ti_2InN$<br>$Zr_2InN$<br>$Ti_3InC_2$ | $Ti_2SnC$<br>$Hf_2SnN$<br>$Zr_2SnC$<br>$Nb_2SnC$<br>$Hf_2SnC$<br>$Lu_2SnC$<br>$V_2SnC$<br>$Sc_2SnC$<br>$Ti_3SnC_2$<br>$Zr_3SnC_2$<br>$Hf_3SnC_2$ | $Ti_2SbP$<br>$Zr_2SbP$<br>$Hf_2SbP$<br>$Nb_2SbC$<br>$Ti_3SbC_2$ | $Zr_2TeP$ |
| **Tl** | **Pb** | **Bi** | **Zn** |
| $Ti_2TlC$<br>$Zr_2TlC$<br>$Hf_2TlC$<br>$Zr_2TlN$ | $Ti_2PbC$<br>$Zr_2PbC$<br>$Hf_2PbC$ | $Nb_2BiC$ | $Ti_2ZnC$<br>$Ti_2ZnN$<br>$V_2ZnC$<br>$Nb_2ZnC$<br>$Ti_3ZnC_2$<br>$Ta_2ZnC$ |
| **Cu** | **Cd** | **Iron triad (Fe, Co, Ni)** | **Precious metals** |
| $Ti_2CuN$<br>$Nb_2CuC$<br>$Ti_4CuN_3$ | $Ti_2CdC$ | $Ta_2FeC$<br>$Nb_2FeC$<br>$Ti_2FeN$<br>$Ta_2CoC$<br>$Nb_2CoC$<br>$Nb_2NiC$<br>$Ta_2NiC$ | $Mo_2AuC$<br>$Ti_2AuN$<br>$Nb_2AuC$<br>$Nb_2PtC$<br>$Ti_3IrC_2$<br>$Ti_3AuC_2$ |

[a] This list is an updated version of list shown in Ref [7].

## Table 4 List of experimentally synthesized MAX alloying phases[a]

| | M site | | A site | X site |
|---|---|---|---|---|
| **211** | $(Ti_{0.02}Cr_{0.98})_2AlC$ | $(Mo_{2/3}Ce_{1/3})_2AlC$ | $Nb_2(Au_{0.5}Al_{0.5})C$ | $Zr_2Se(B_{0.94}Se_{0.06})$ |
| | $(Ti_{0.25}Cr_{0.75})_2AlC$ | $(Mo_{2/3}Pr_{1/3})_2AlC$ | $Nb_2(Ag_{0.3}Sb_{0.4}Al_{0.3})C$ | $Zr_2Se(B_{0.93}Se_{0.07})$ |
| | $(V_{0.65}Ta_{0.35})_2AlC$ | $(Mo_{2/3}Nd_{1/3})_2AlC$ | $Nb_2(Pd_{0.5}Sn_{0.5})C$ | $Zr_2Se(B_{0.89}Se_{0.11})$ |
| | $(Ti_{0.4}Ta_{0.6})_2AlC$ | $(Mo_{2/3}Sm_{1/3})_2AlC$ | $Nb_2(Pt_{0.6}Al_{0.4})C$ | $Zr_2Se(B_{0.83}Se_{0.17})$ |
| | $(Sc_{0.33}Nb_{0.67})_2AlC$ | $(Mo_{2/3}Tb_{1/3})_2AlC$ | $Nb_2(Rh_{0.2}Sn_{0.4}Al_{0.4})C$ | $Zr_2Se(B_{0.64}Se_{0.36})$ |
| | $(V_{0.96}Mn_{0.04})_2AlC$ | $(Mo_{2/3}Dy_{1/3})_2AlC$ | $Nb_2(Ge_{0.8}Al_{0.2})C$ | $Zr_2Se(B_{0.40}Se_{0.60})$ |
| | $(Mo_{0.5}Mn_{0.5})_2GaC$ | $(Mo_{2/3}Ho_{1/3})_2AlC$ | $Zr_2(Al_{0.42}Bi_{0.58})C$ | $Zr_2Se(B_{0.03}Se_{0.97})$ |
| | $(Cr_{0.5}Mn_{0.5})_2GaC$ | $(Mo_{2/3}Er_{1/3})_2AlC$ | $Cr_2(Al_{0.97}Si_{0.03})C$ | |
| | $(Ti_{0.5}Zr_{0.5})_2InC$ | $(Mo_{2/3}Tm_{1/3})_2AlC$ | $Zr_2(Al_{0.3}Sb_{0.7})C$ | |
| | $(Ti_{0.5}Hf_{0.5})_2InC$ | $(Mo_{2/3}Lu_{1/3})_2AlC$ | $Zr_2(Al_{0.35}Pb_{0.65})C$ | |
| | $(Ti_{0.47}Hf_{0.53})_2InC$ | $(Mo_{2/3}Sc_{1/3})_2GaC$ | $Mo_2(Ga_{0.33}Fe_{0.5}Au_{0.16})C$ | |
| | $(Ti_{0.5}V_{0.5})_2GeC$ | $(Mo_{2/3}Y_{1/3})_2GaC$ | | |
| | $(Cr_{0.5}Mn_{0.5})_2AuC$ | $(Cr_{2/3}Sc_{1/3})_2GaC$ | | |
| | $(Mo_{2/3}Sc_{1/3})_2AlC$ | $(Mn_{2/3}Sc_{1/3})_2GaC$ | | |
| | $(Mo_{2/3}Y_{1/3})_2AlC$ | $(Mo_{2/3}Gd_{1/3})_2GaC$ | | |
| | $(Cr_{2/3}Sc_{1/3})_2AlC$ | $(Mo_{2/3}Tb_{1/3})_2GaC$ | | |
| | $(Cr_{2/3}Y_{1/3})_2AlC$ | $(Mo_{2/3}Dy_{1/3})_2GaC$ | | |
| | $(W_{2/3}Sc_{1/3})_2AlC$ | $(Mo_{2/3}Ho_{1/3})_2GaC$ | | |
| | $(W_{2/3}Y_{1/3})_2AlC$ | $(Mo_{2/3}Er_{1/3})_2GaC$ | | |
| | $(Cr_{2/3}Zr_{1/3})_2AlC$ | $(Mo_{2/3}Tm_{1/3})_2GaC$ | | |
| | $(V_{2/3}Zr_{1/3})_2AlC$ | $(Mo_{2/3}Yb_{1/3})_2GaC$ | | |
| | $(V_{2/3}Sc_{1/3})_2AlC$ | $(Mo_{2/3}Lu_{1/3})_2GaC$ | | |
| | $(V_{0.75}Cr_{0.25})_2AlC$ | $(Cr_{0.25}Ti_{0.75})_2AlC$ | | |
| | $(V_{0.25}Cr_{0.75})_2AlC$ | $(Nb_{2/3}Sc_{1/3})_2AlC$ | | |
| | $(V_{0.5}Cr_{0.5})_2AlC$ | $(Ti_{0.25}V_{0.75})_2AlC$ | | |
| | $(Cr_{0.5}V_{0.5})_2GeC$ | $(Ti_{0.5}V_{0.5})_2AlC$ | | |
| | $(Mo_{2/3}Gd_{1/3})_2AlC$ | $(Ti_{0.75}V_{0.25})_2AlC$ | | |
| **312** | $(Cr_{2/3}Ti_{1/3})_3AlC_2$ | | $Ti_3(Sb_{0.5}Sn_{0.5})CN$ | $Ti_3SbCN$ |
| | $(Cr_{2/3}V_{1/3})_3AlC_2$ | | $Ti_3(Sb_{0.5}Sn_{0.5})C_2$ | $Ti_3AlCN$ |
| | $(Mo_{2/3}Ti_{1/3})_3AlC_2$ | | $Ti_3(Cd_{0.5}Zn_{0.5})C_2$ | $Ti_3SnCN$ |
| | $(Mo_{2/3}Sc_{1/3})_3AlC_2$ | | $Ti_3(Al_{0.8}Sn_{0.2})C_2$ | $Ti_3GaCN$ |
| | | | $Ta_3(Al_{0.96}Sn_{0.04})C_2$ | |
| **413** | $(Nb_{0.5}V_{0.5})_4AlC_3$ | | | |
| | $(Nb_{0.8}Ti_{0.2})_4AlC_3$ | | | |
| | $(Nb_{0.8}Zr_{0.2})_4AlC_3$ | | | |
| | $(Cr_{5/8}Ti_{3/8})_4AlC_3$ | | | |
| | $(V_{0.5}Cr_{0.5})_4AlC_3$ | | | |
| | $(Mo_{0.5}Ti_{0.5})_4AlC_3$ | | | |

[a] This list is an updated version of lists shown in Ref [7] and [33].

**Table 5 List of non-formable MAX phase screened via *ab initio* calculations [26].**

| 211 | | 312 | | 413 | |
|---|---|---|---|---|---|
| $Cr_2InC$ | $Mo_2GaN$ | $Cr_3InC_2$ | $Mo_3InN_2$ | $Cr_4GaC_3$ | $Mo_4AsC_3$ |
| $Cr_2TlC$ | $Mo_2InN$ | $Cr_3TlC_2$ | $Mo_3TlN_2$ | $Cr_4InC_3$ | $Mo_4SC_3$ |
| $Cr_2SnC$ | $Mo_2TlN$ | $Cr_3GeC_2$ | $Mo_3SiN_2$ | $Cr_4TlC_3$ | $V_4PbN_3$ |
| $Cr_2PbC$ | $Mo_2SiN$ | $Cr_3SnC_2$ | $Mo_3GeN_2$ | $Cr_4SiC_3$ | $Ta_4TlN_3$ |
| $Cr_2SC$ | $Mo_2GeN$ | $Cr_3PbC_2$ | $Mo_3SnN_2$ | $Cr_4GeC_3$ | $Ta_4PbN_3$ |
| $Mo_2InC$ | $Mo_2PbN$ | $Cr_3AsC_2$ | $Mo_3PbN_2$ | $Cr_4SnC_3$ | $Cr_4SiN_3$ |
| $Mo_2TlC$ | $Mo_2PN$ | $Mo_3GaC_2$ | $Mo_3PN_2$ | $Cr_4PbC_3$ | $Cr_4PbN_3$ |
| $Mo_2SnC$ | $Mo_2AsN$ | $Mo_3InC_2$ | $Mo_3AsN_2$ | $Cr_4PC_3$ | $Cr_4PN_3$ |
| $Mo_2PbC$ | | $Mo_3TlC_2$ | | $Cr_4AsC_3$ | $Cr_4SN_3$ |
| $Mo_2SC$ | | $Mo_3SiC_2$ | | $Cr_4SC_3$ | $Mo_4GaN_3$ |
| $V_2SN$ | | $Mo_3GeC_2$ | | $Mo_4AlC_3$ | $Mo_4InN_3$ |
| $Nb_2GaN$ | | $Mo_3SnC_2$ | | $Mo_4GaC_3$ | $Mo_4TlN_3$ |
| $Nb_2SN$ | | $Mo_3PbC_2$ | | $Mo_4InC_3$ | $Mo_4SiN_3$ |
| $Ta_2InN$ | | $Mo_3AsC_2$ | | $Mo_4TlC_3$ | $Mo_4GeN_3$ |
| $Ta_2GeN$ | | $Mo_3SC_2$ | | $Mo_4SiC_3$ | $Mo_4SnN_3$ |
| $Cr_2TlN$ | | $Ta_3PbN_2$ | | $Mo_4GeC_3$ | $Mo_4PbN_3$ |
| $Cr_2PbN$ | | $Cr_3PN_2$ | | $Mo_4SnC_3$ | $Mo_4PN_3$ |
| $Cr_2PN$ | | $Mo_3AlN_2$ | | $Mo_4PbC_3$ | $Mo_4AsN_3$ |
| $Mo_2AlN$ | | $Mo_3GaN_2$ | | $Mo_4PC_3$ | $Mo_4SN_3$ |

## 4 Results and discussion

### 4.1 The structure maps construction for MAX phases formability

Firstly, the quest for generality raises the inquiry of whether the *geometrical* and *electron concentration* factors proposed in Ref. [6] could be extended to a broad family of $M_{n+1}AX_n$ phases; and Fig. 3 (a)-(c) show the constructed structure maps through *geometrical* and *electron concentration* factors proposed in Ref. [6] for 211-MAX phases, 312-MAX phases and (c) 413-MAX phases respectively, where the outliers are marked with labels of corresponding compositions. It is shown that, with fidelity of 92.1%, the applications of the *geometrical* and *electron concentration* factors proposed *in prior* to broad MAX-phase members are feasible. Nevertheless, it is expected to further enhance the fidelity through upgrading the expressions of *geometrical* factor, by taking the X-site elements dimension, as well as different *n* values, into considerations.

The renewal structure maps of MAX phases formability are constructed from Eqn. (2) and Eqn. (6), which are shown in Fig. 4-6 for 211-, 312- and 413-MAX phases respectively (The structure maps with full set of labels are shown in S1-S3 in Supplement

Information). It can be found that, only with a few exceptions, the points representing formable MAX phases and those of non-formable ones are located within two separable regions; and successful formability/non-formability separation is achieved with a fidelity of 95.5%.

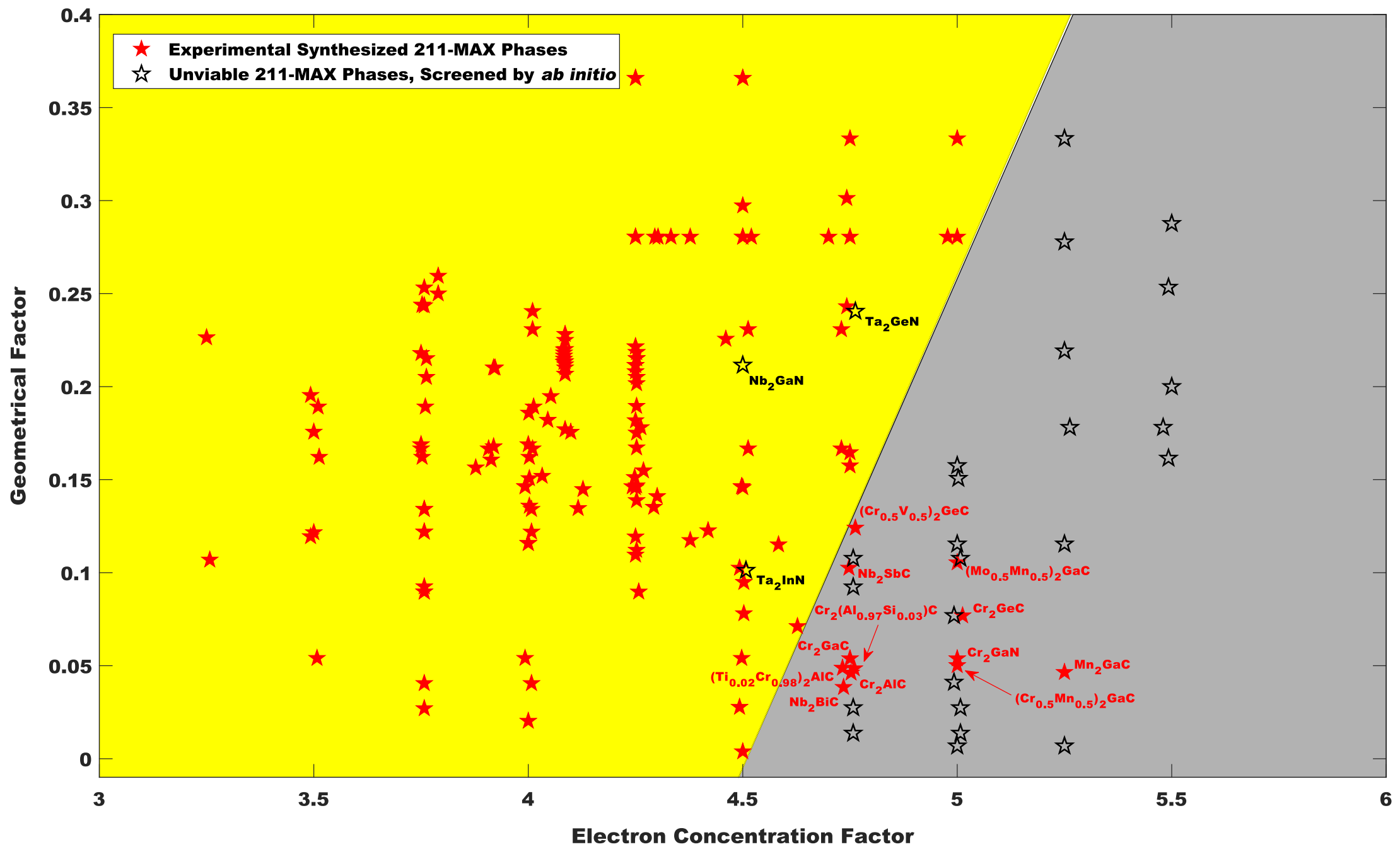

(a)

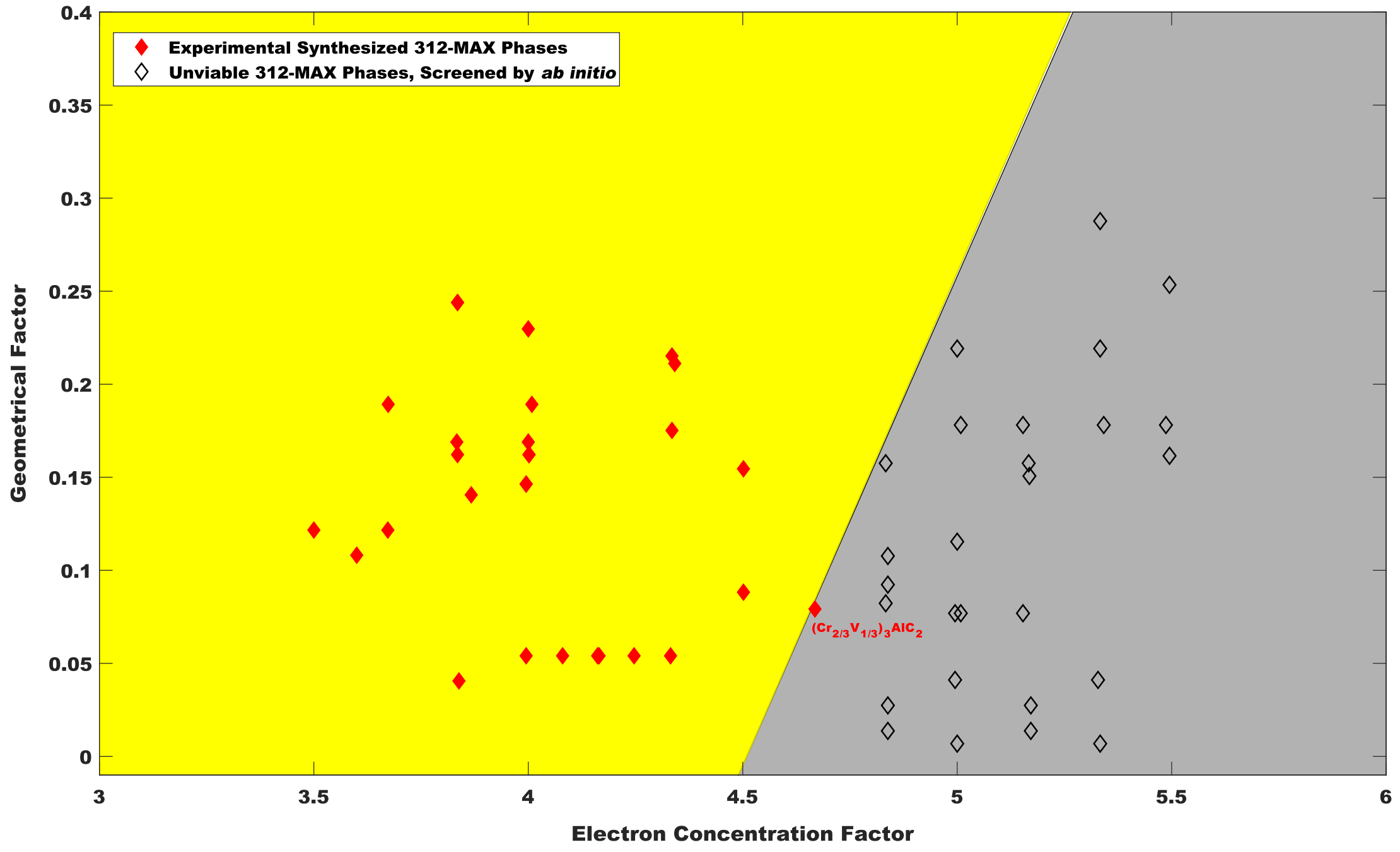

(b)

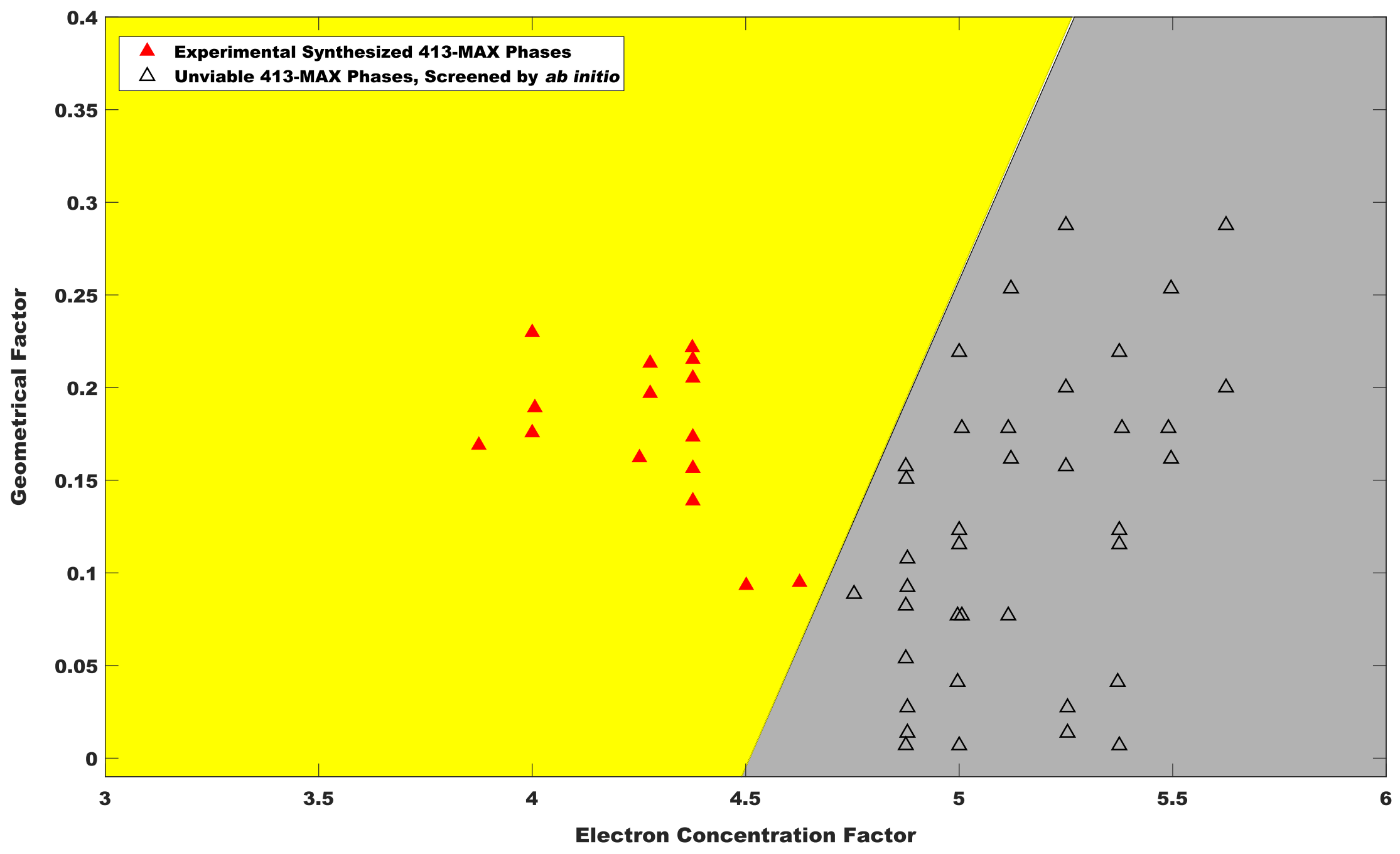

(c)

Fig. 3 The structure maps constructed *via* geometrical and electron concentration factors proposed in Ref. [6] for (a) 211-MAX phases; (b) 312-MAX phases; and (c) 413-MAX phases

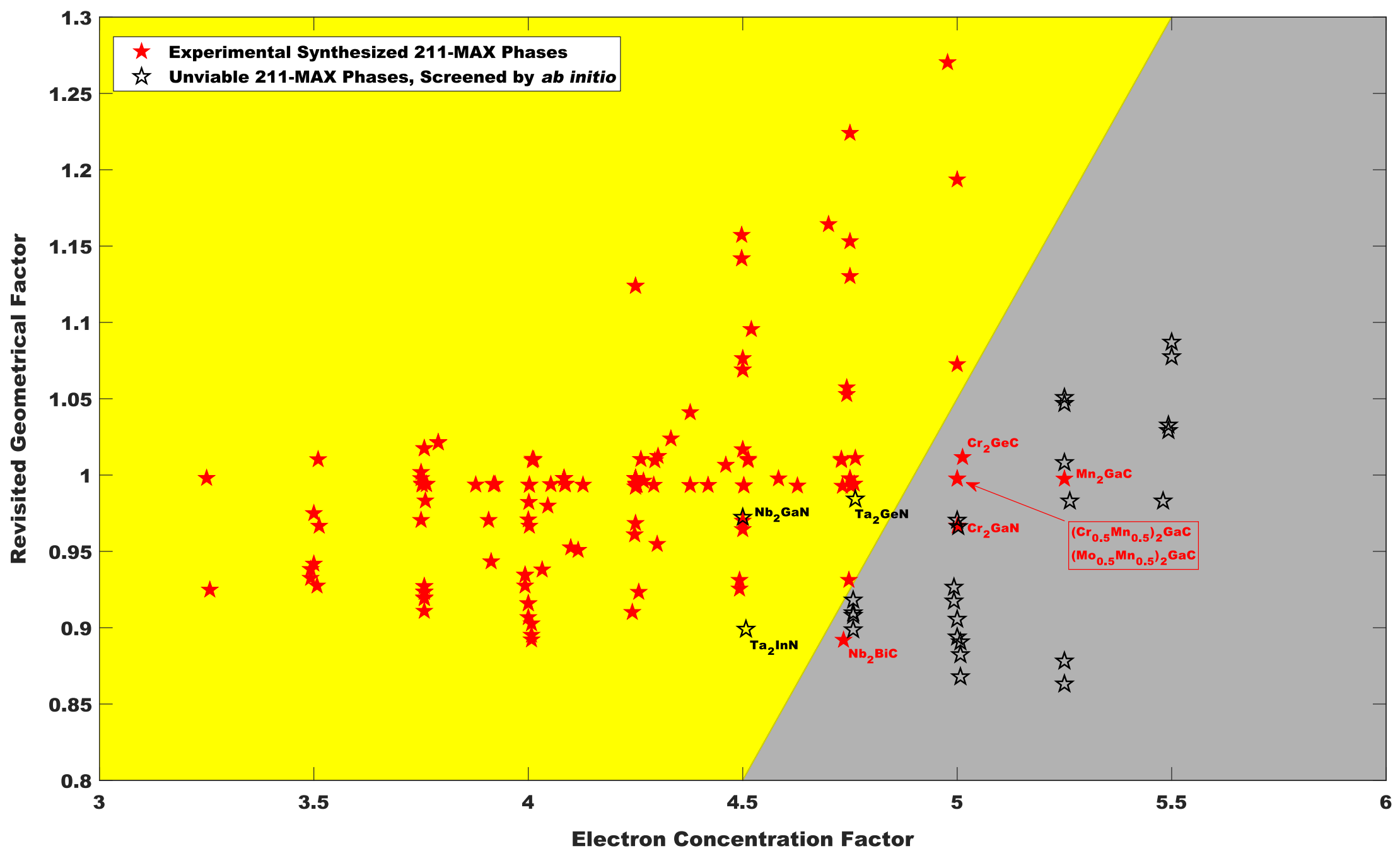

Fig. 4 The structure maps constructed *via* revisited geometrical and electron concentration factors for (a) 211-MAX phases (ternary) and (b) 211-MAX alloying phases.

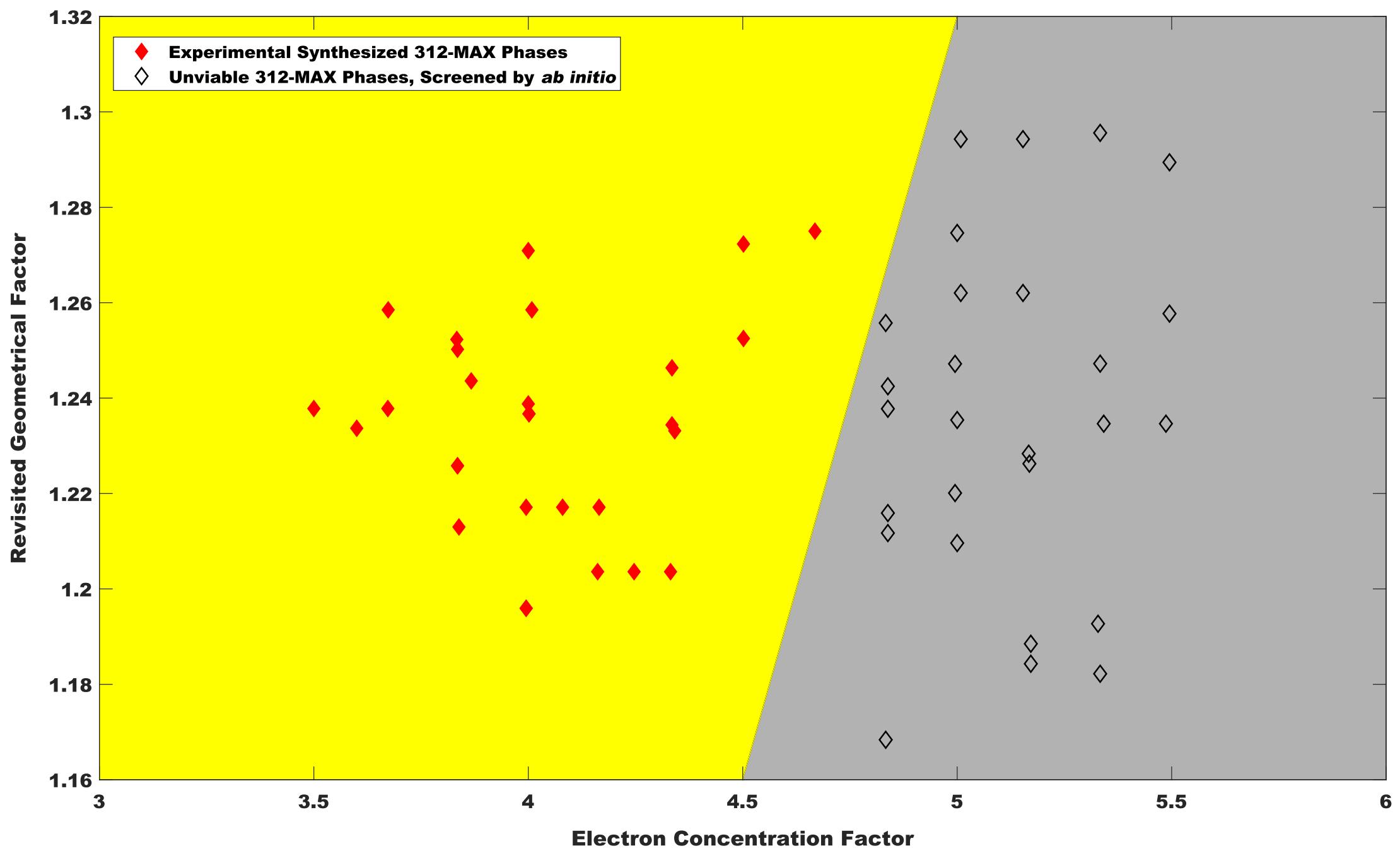

Fig. 5 The structure maps constructed *via* revisited geometrical and electron concentration factors for 312-MAX phases.

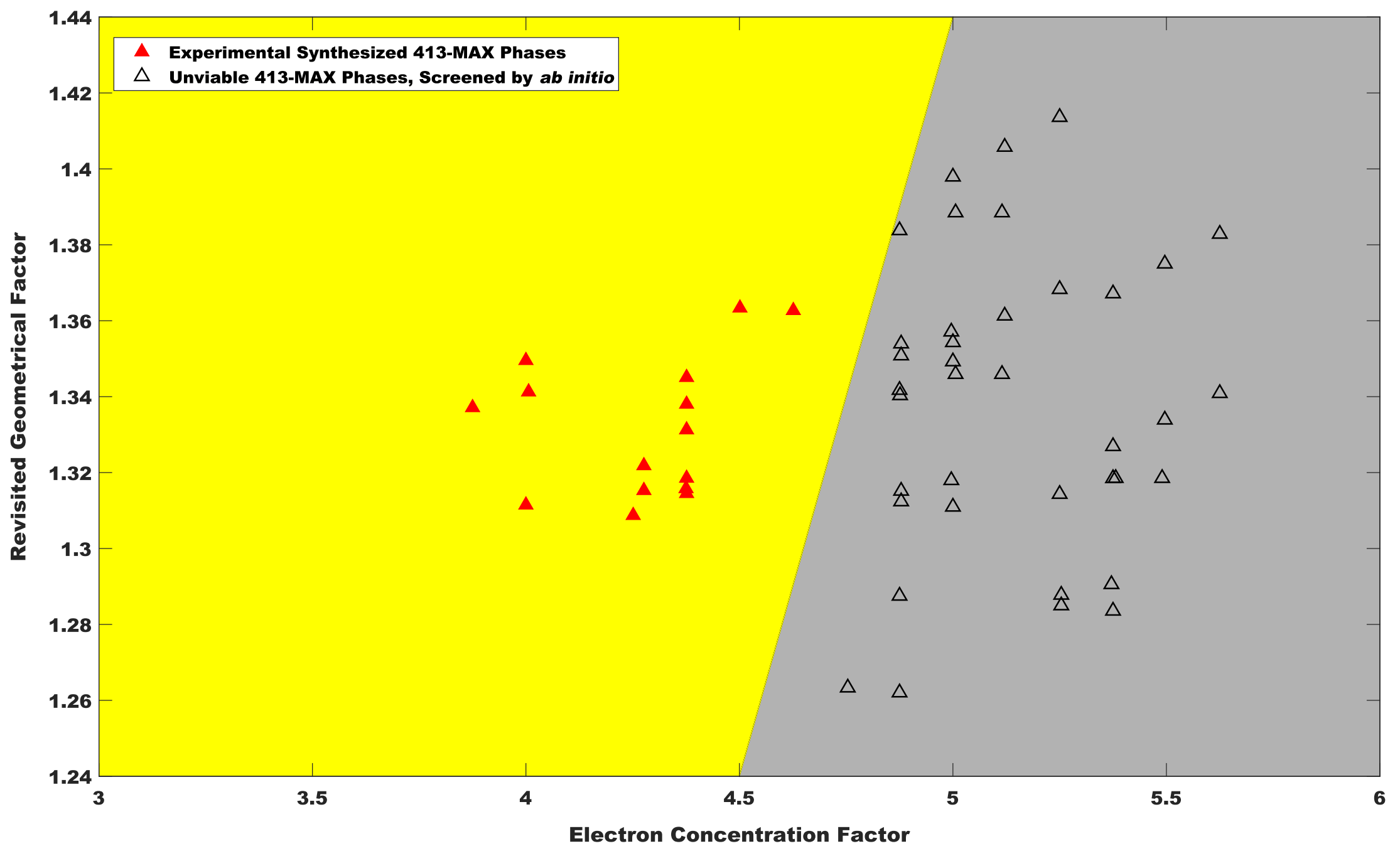

Fig. 6 The structure maps constructed *via* revisited geometrical and electron concentration factors for 413-MAX phases.

### 4.2 The MAX phases with high electron concentration elements and the effects of geometrical factor and alloying

In Fig. 4, it is shown that several synthesized Cr-, Mn- and Nb- containing 211-MAX phases (*i.e.* $Cr_2GeC$, $Cr_2GaN$, $Mn_2GaC$, $Nb_2BiC$, $(Mo_{0.5}Mn_{0.5})_2GaC$, $(Cr_{0.5}Mn_{0.5})_2GaC$) locate within the non-formable region. In Barsoum's monograph, the phases associated with high electron concentrations are fully discussed and referred as anomalous or borderline-stable compounds; where the phases of $Cr_2GeC$, $Cr_2AlC$ and $V_2AsC$ are taken as examples [2].

However, the geometrical factor plays an important role in offsetting the destabilizing effect of electron concentration factor, when taking the MAX phases containing Mo- as examples. The formability/non-formability of Mo- containing phases are distinctly separated, where the formability of $Mo_2GaC$ could be attributed to its geometrical factor. In addition, due to the contributions from geometrical factors of $Cr_2AlC$ and $V_2AsC$, these two mentioned borderline-stable compounds are properly classified as formable ones.

Further, it is shown that the alloying also imposes stabilizing effects *via* tuning the electron concentrations. For example, the compounds of $Cr_3AlC_2$ and $Mo_3AlC_2$ do not exist [78], but $(Cr_{2/3}Ti_{1/3})_3AlC_2$ [79], $(Cr_{2/3}V_{1/3})_3AlC_2$ [80], $(Mo_{2/3}Ti_{1/3})_3AlC_2$ [81] and $(Mo_{2/3}Sc_{1/3})_3AlC_2$ [82] have been successfully synthesized. In addition, while the $Mo_4AlC_3$ phase is screened out as non-formable one, $(Mo_{0.5}Ti_{0.5})_4AlC_3$ has been derived experimentally [82].

### 4.3 The validity of *electron concentration* and *geometrical factors*

From the results depicted in Fig. 4-6, it is found that in 2D *electron concentration factor* vs. *geometrical factor* maps, the formable and non-formable MAX phases tend to cluster in separated regions. Accordingly, the distinguishing power for formable and non-formable of MAX phases from the compounding effect of *electron concentration* and *geometrical factor* can be fully acknowledged.

However, the *geometrical factor* suffers from relying heavily on empirical atomic radii. It is well known that the atomic radii in compounding state is not precisely defined and that even within the same definition there are variations reflecting the coordination and local chemistry [83-87]. Due to these uncertainties transfer to the *intercalation* and *octahedral* factors, the predictive power for future MAX phases might be weaken. To overcome these limitations, precise atomic radii determined from first principles should be introduced.

For *electron concentration* factor, the number of valence electrons is applied to M-site elements, by considering the covalent characteristic of M-X bonding. However, the heterodesmic characteristics possessed by M-site elements (where M-X bonding is covalent, and M-A bonding is metallic) requires further elaborations for the choice of electron concentration parameters [65].

### 4.4 The gap in the capability of distinction for the chemical order and disorder

The families of MAX phases have significantly expanded within last few years, in particular for the quaternary and even quinary phases are derived upon alloying. In 2014, one chemically ordered MAX phase, *i.e.* $Cr_2TiAlC_2$, was reported [79], which displays out-of-plane order from alternating layers composed of one M element only. Other out-of-plane ordered MAX phases then followed soon [80-82]. In 2017, MAX phases with in-plane ordering of M-site elements of a 2:1 ratio were discovered [88, 89]. Further, the expansions of MAX phase members have also been manifested through complete or partial substation on the A-site elements [8, 90-96].

It is well acknowledged that, the understanding of why certain metal combinations would lead to the formation of i-MAX, o-MAX or solid solution phases is a fundamental question. Recently, Dahlqvist and Rosen [33] has carried out an assessment for identifying the formation rules defining the preference for chemical order (i-MAX phase) or disorder (solid solution), by employing energy difference calculations on 2702 unique compositions. In contrast, the present work mainly focuses on making the robust and efficient associations between the chemical formular of $M_{n+1}AX_n$ to the structure type of MAX phases; and another classification theme should be built for further distinguishing the phases with chemical order and disorder.

### 4.5 What is the generality belongs to: the *proposed factors* or the *developed maps*?

In recent times, we have noticed Carlsson *et al.* [30] have challenged the use of structure mapping for identifying formable MAX phases [6, 15]. We are fully appreciated the critical assessments imposed on our developed strategy; however, several notions have to be illuminated here. Firstly, the developed structure map is aiming for mapping MAX phase structures to their compositions with chemical formular expression of $M_{n+1}AX_n$, on which the lines that distinguish the formable regions are defined. While in Carlsson *et al.*'s work, they directly applied the defined lines on the case of MAB phases, with chemical formulas of MAB, $M_2AB_2$, $M_3AB_4$, $M_4AB_4$ and $M_4AB_6$, all of which have different chemical formula style from $M_{n+1}AX_n$. Therefore, the developed structure map in Ref. [6], as well as the defined lines to distinguish the formable regions, cannot applied directly. Secondly, Carlsson *et al.* have put all the orthorhombic and hexagonal MAB phases (that with different chemical formular styles) together for mapping the phase stable region. It has to be emphasized that compounds with different chemical formular styles cannot be put together, and separate mappings have to be conducted for each chemical formular style. Thirdly, the Fig. S15-S19 in Carlsson *et al.*'s work shows the mappings for each considered composition (*i.e.* MAB, $M_2AB_2$, $M_3AB_4$, $M_4AB_4$ and $M_4AB_6$) separately, where the visible separations can be seen. Therefrom the lines for distinguishing the formable regions can be derived accordingly for each case. Based on above analysis, there are the generalities that proposed factors have, but not belong to the developed maps that are specific for $M_{n+1}AX_n$ phases.

## 5 Conclusions

As one kind of heterodesmic compounds, MAX phases possess chemical bonding of metallic, ionic, covalent, or a changing mixture of those. Conventionally, the explorations of the uncharted territories of MAX phases, with respect to their composition and structure, demand conducting of trail-and-error experimentations or analytical solutions of Schrödinger equations. These tactics have shown their values in determining given MAX phases; yet, the intensive characteristics restrict their efficiencies for future candidate investigations. In this work, we have employed phenomenological structure mapping strategy to develop a classification scheme, with a fidelity of 95.5%, for MAX phases formability/non-formability separation. More generally, current findings suggest that the proposed genomic blueprints, that is *geometrical* and *electron concentration* factors, could serve as a powerful tool to tackle the factorial complexity of combinatorial MAX phases design.

**Acknowledgement**

The authors acknowledge the financial support of Pioneer and Leading Goose R&D Program of Zhejiang (No. 2022C01236), National Natural Science Foundation of China (No. 52250005, U2004212), and Ten-Thousand Talents Plan of Zhejiang Province (No. 2022R51007).

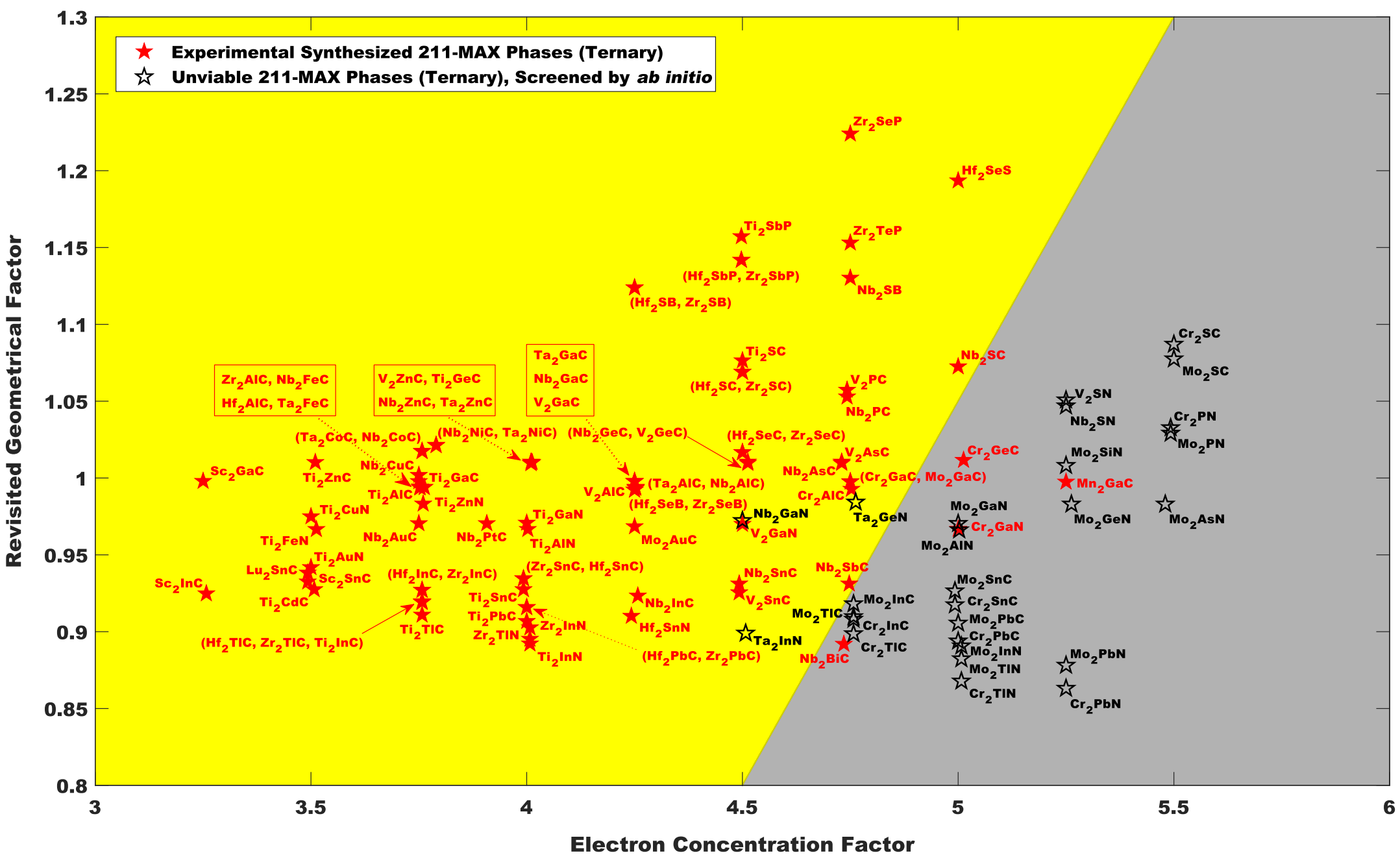

(a)

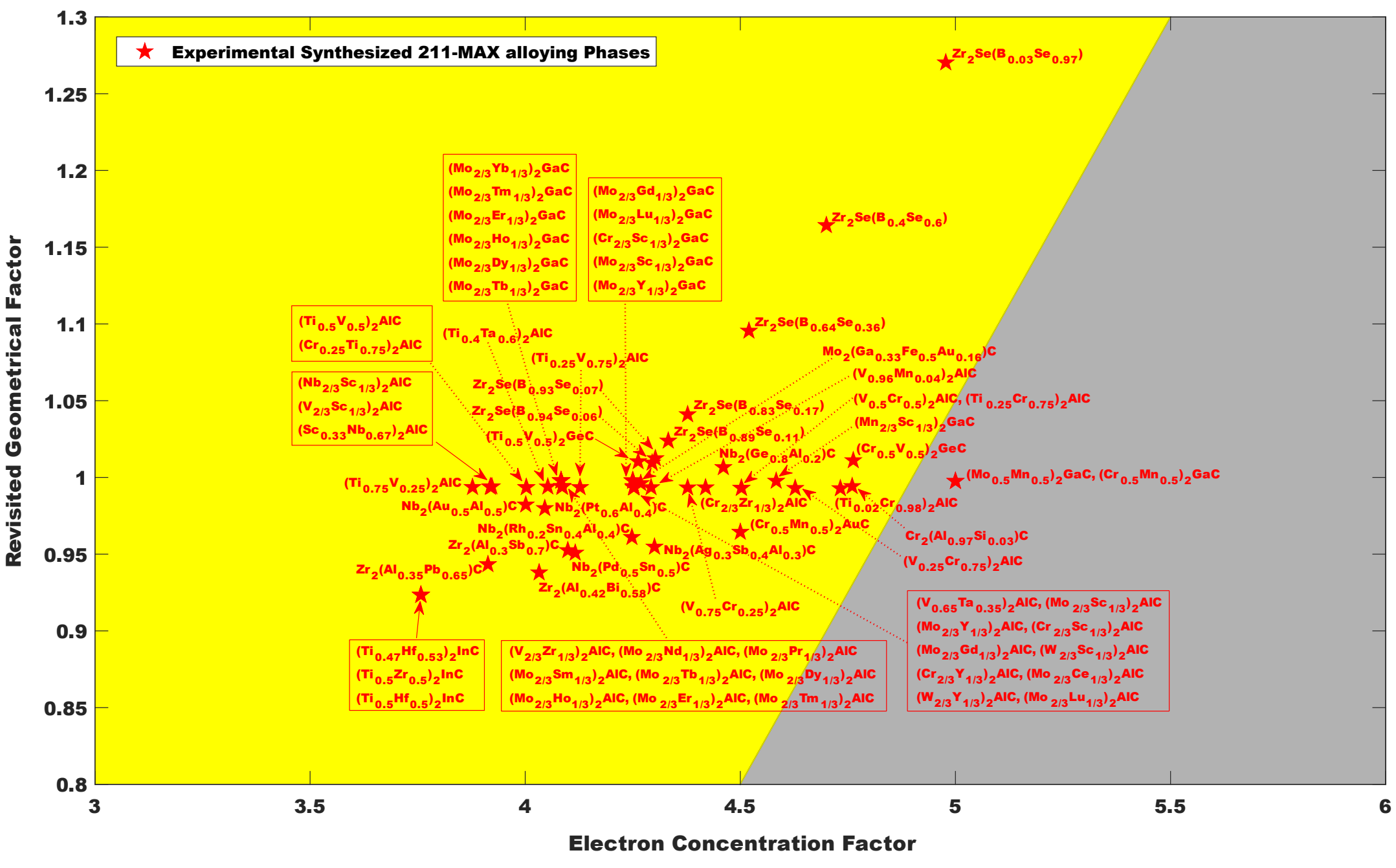

(b)

S1 The structure maps constructed *via* revisited geometrical and electron concentration factors for (a) 211-MAX phases (ternary) and (b) 211-MAX alloying phases (with full set of labels).

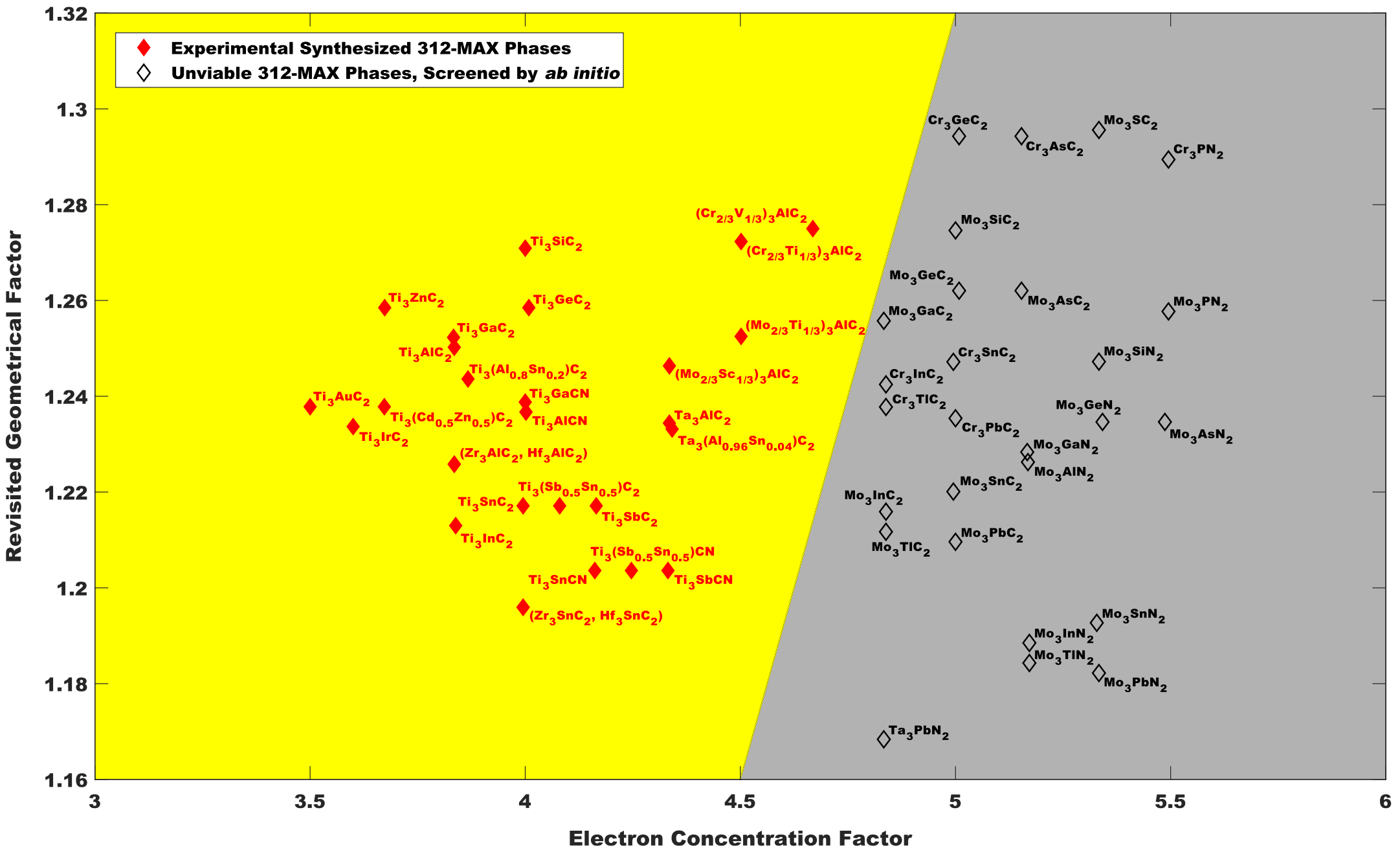

S2 The structure maps constructed *via* revisited geometrical and electron concentration factors for 312-MAX phases (with full set of labels).

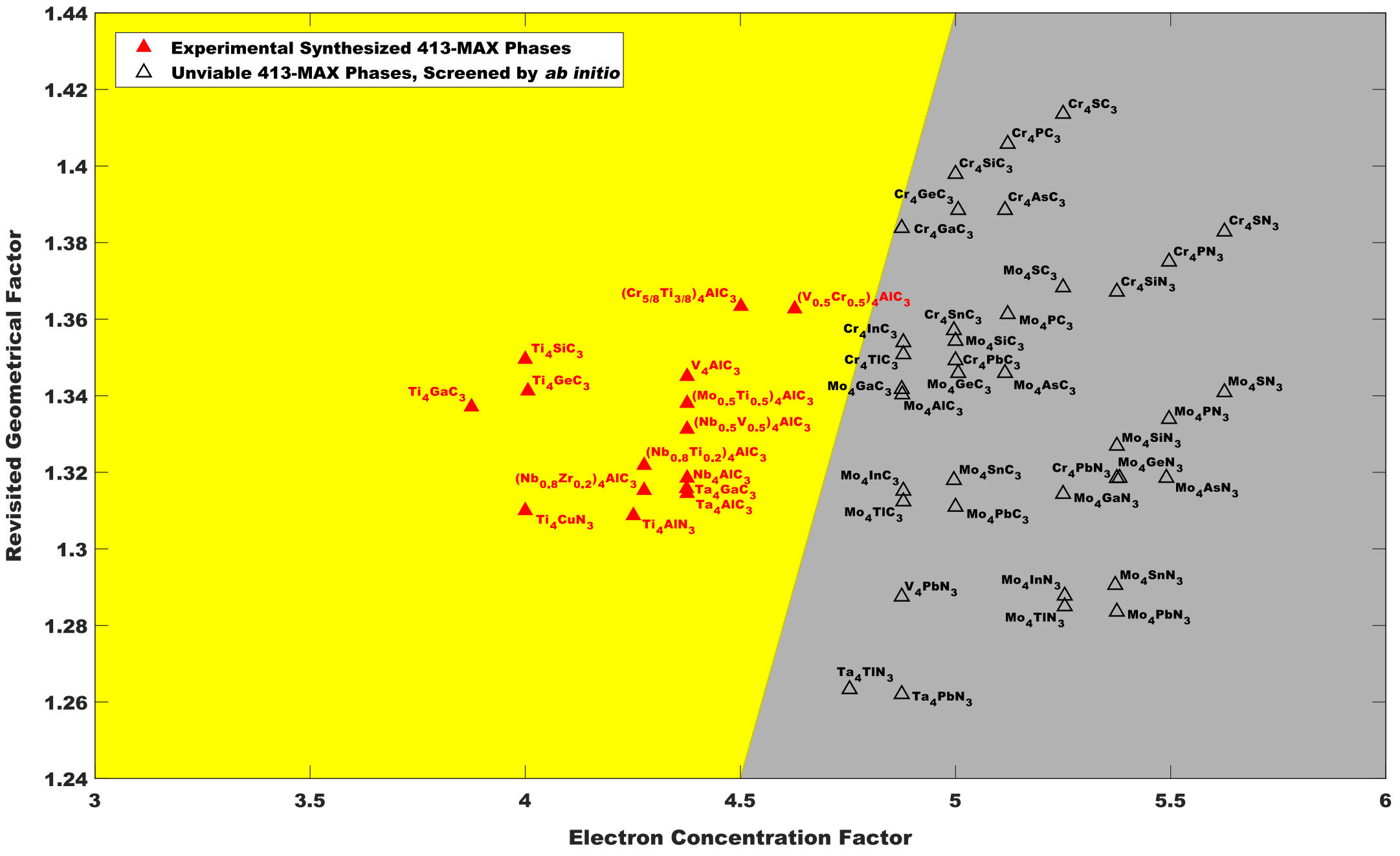

S3 The structure maps constructed *via* revisited geometrical and electron concentration factors for 413-MAX phases (with full set of labels).